## Phase diagram and Raman Imaging of Grain Growth Mechanisms in Highly Textured Pb(Mg$_{1/3}$Nb$_{2/3}$)O$_3$-PbTiO$_3$ Piezoelectric Ceramics

Mai Pham Thi[1], Gregory March[2], Philippe Colomban [2*]

[1]Thales Research & Technology France, Orsay, France

[2]Nanophases and Heterogeneous Solids Group,

Ladir, UMR 7075 CNRS & Université Pierre & Marie Curie,

2 rue Henry-Dunant, 94320 Thiais, France

**Abstract**

Pb(Mg$_{1/3}$Nb$_{2/3}$)O$_3$-PbTiO$_3$ solid solution (PMN$_{(1-x)}$-PT$_x$) ceramics with 0 <x<1 have been prepared by homo-epitaxial templated grain growth (HTGG) using cubic PMN-PT single crystal seeds as templates and nanoparticles for thes ceramic matrix. Samples were prepared using the doctor Blade method and sintered in a controlled atmosphere. They were then analysed by DTA, TGA, optical/scanning electron microscopy, Raman spectroscopy and imaging. Relationships between Raman parameters such as band component or area versus composition (x) are proposed. Wavenumber and peak area shifts from the single crystal seed centre to periphery growth are assigned to compositional change. Representative medium to highly textured ceramics with Lotgering factors of ~0.7 and 0.9 were sintered at 1150°C and 1200°C, respectively, and studied by direct and smart Raman imaging. The Raman peak centre of gravity depends on the solid solution composition whereas peak intensity is correlated to the unit-cell distorsion and related short-range structure. Two different growth mechanisms are observed: below ~1200°C, sharp corner and straight edged single crystal seeds develop through a liquid I phase at the crystal/matrix interface; at higher temperatures, the matrix is consumed and crystal growth develops in contact with a liquid II phase, through the help of the high PbO partial pressure. Smart Raman imaging shows that the final composition and structure is very close to that of the matrix. Formation of short-range ordered B/B' domains is expected. A phase diagram is proposed.

* corresponding author : colomban@glvt-cnrs.fr

fax : 33 1 4978 1118

# 1. INTRODUCTION

The piezoelectric performance of relaxor single crystals provides excellent opportunities for improvements in medical ultrasonic imaging, sonar applications, active damping and high strain actuation, if the costs can be reduced. PMN(PZN)-PT single crystals exhibit unique electromechanical properties when poled along the non-polar <100> direction. This is related to the fact that single crystal domains easily oriente during poling so that the maximum response is obtained. A longitudinal coupling coefficient $k_{33}$ of 90% and an exceptional piezoelectric coefficient $d_{33} > 2000$ pC/N can be obtained (1-2).. While expensive crystal growth techniques are advancing, it is of great practical importance to develop an alternative low-cost production method based on strongly oriented or textured ceramics.

$Pb(Mg_{1/3}Nb_{2/3})O_3$-$PbTiO_3$ ($PMN_{(1-x)}$-$PT_x$) solid solution offers a large set of ferroelectric/piezoelectric properties. Ceramics elaborated by solid-state sintering (x = 0.35) exhibit an electro-mechanical coupling efficiency of $k_{33} = 70\%$, $\varepsilon \sim 4000$ and a longitudinal $d_{33}$ piezoelectric coefficient of 500 pC/N (3). Due to the random crystallographic orientation of the bulk ceramic, engineering of ferroelectric domains cannot be achieved and one gets an average of the directionally dependent physical properties.

Ceramic texturing via Templated Grain Growth (TGG) consists in orientating a dispersion of large anisotropic template particles in a green state fine particle matrix. During sintering, seed growth consumes the surrounding matrix grains and creates a crystallographically oriented ceramic. TGG process efficiency has been demonstrated in a wide variety of systems, including $Al_2O_3$, mullite, $Bi_4Ti_3O_{12}$ and $Sr_2Nb_2O_7$ (4,5). To date, textured $Pb(Mg_{1/3}Nb_{2/3})O_3$-$PbTiO_3$ ceramics have been obtained by Reactive Templated Grain Growth (RTGG) using Ba/SrTiO$_3$ seeds (6). The problem with heterogeneous seeding is the residual presence of templates (6,7). The strain-electric field hysteresis is then larger than in random ceramics while the electro-mechanical properties measured by low-field resonance are still much smaller than observed in single crystals (8). Un-reacted $BaTiO_3$ templates exhibiting low piezoelectric properties reduce the effective properties of textured samples.

We demonstrated recently that textured PMN-PT ceramics could be prepared by homo-epitaxy template grain growth using PMN-PT templates (HTGG) (9-10). Alternative method using an initially cracked single crystal seed plate was reported (11,12). Any matrix composition can be achieved but crystal seed composition must match with the phase diagram and therefore are limited in number. In this work we first analysed single crystal seeds and pure ceramics of different compositions (0<x<1) by Raman spectroscopy, in order to establish a relationship between Raman parameters and composition distribution. Our goal is to check the validity of Smart Raman Imaging as a control method for textured ceramics. We then studied the grain growth process and the composition homogenisation for two different seed/matrix couples, representative of the first and final growing process. Topological, chemical and crystallographic transformations are discussed in the light of x-ray diffraction, TGA, DTA and Raman imaging results.

## 2. EXPERIMENTAL PROCEDURE

**Samples**

*Matrix Powder*

PMN-PT nanoparticles were synthesized using the columbite method (11) and milled by continuous attrition milling with micro-beads.

*Seeds*

PMN-PT cubic shaped (x~ 0.25) single crystals were prepared using the flux method (10). Crystals with size close to 20-50 µm were selected by sieving.

*Ceramics*

PMN-PT nano-powders and a few weight percent of PMN-PT single crystal templates were mixed with an organic binder to obtain a homogenous slurry. Two weight percent of PbO was added. The mixture was tape-cast using a doctor blade. Green tapes of 100-150 µm thick were dried, laminated and cut into disk (diameter = 16 mm, 0.5 to 3 mm thick). Ceramic sintering was made under $O_2$ flow between 1150 and 1200°C for different durations in a PbO-rich atmosphere controlled using crucibles packed in a PbO-rich powder mixture..

**Techniques**

*Optical microscopy*

Microstructures were observed on as-sintered and polished surfaces using an Olympus optical microscope (up to x1000 magnification).

*TGA and DTA*

Measurements were recorded on a TA instrument apparatus. All experiments were performed by heating/cooling cycles at a rate of 20°C/mn in a Pt open-crucible under $O_2$ flow. The rather high heating/cooling rate prevents excessive volatilisation of the PbO.

Density was measured using the Archimedes method.

*Raman microspectrometry*

An "XY" spectrograph (Dilor, France) equipped with a double monochromator filter and a back-illuminated, liquid nitrogen-cooled, 2000 x 256 pixels CCD detector (Spex, Jobin-Yvon–Horiba Company) was used for recording down to ca. 10 cm$^{-1}$ with a 647.1 nm excitation line (slit = 80 μm). A514.53 nm excitation was also used to obtain a large, one shot, spectral window. Various objectives from MSPlan, Japan (numerical aperture = 0.50-0.80; magnification = 10, 50 and 100) were used; the total magnifications are 100, 500 and 1000 and the confocal hole aperture was ~ 100 μm. Rare spikes (cosmic ray) and some plasma lines were removed from the spectra. The laser spot waist was 1 μm with the x100 objective.

*Raman imaging*: In order to map the growth seed distribution, 2000 Raman spectra were recorded over a large area (typically: 200x700 μm$^2$) using a computer-controlled XY table (12). During the measurement all moving parts are fixed and a mercury lamp emission is used to control the grating position. This procedure guaranties a reproducibility of +- one pixel, i.e. better than +- 1cm$^{-1}$ for the used excitation wavelengths.

*X-ray diffraction*

Patterns were obtained using a Philips Instrument. The degree of <001> orientation was estimated from the XRD patterns using the Lotgering method (13). The Lotgering factor (f) is defined as the fraction of area textured with the crystallographic plane of interest. The degree of <001>-texturing for the PMN-PT ceramics was calculated with the Lotgering formula:

$$f_{(00\lambda)} \; = \; P_{(00\lambda)} \; P_0 \; / \; (1 - P_0) \qquad (1)$$

$$P_{(00\lambda)} = \Sigma \; I_{(00\lambda)} \; / \; \Sigma(hkl) \; ; \; P_0 = \Sigma I_{0(00\lambda)} \; / \; \Sigma I_0(hkl) \qquad (2)$$

$\Sigma I_{(00\lambda)}$ is the sum for the textured sampl of all {00l} XRD peaks intensities; $\Sigma I_{(hkl)}$ is calculated the same way with all {hkl} peaks. $\Sigma I_{0(00\lambda)}$ and $\Sigma I_{0(hkl)}$ are the XRD peak intensities for the randomly oriented sample.

*Peak fitting and data processing*

In undertaking a curve fit of a single Raman spectrum, the linear baseline was first subtracted using LabSpec® software (Dilor, France). An Origin peak-fitting module (Microcal® Software, Inc., USA) was used to calculate the integral area, the bandwidth and the centre of gravity for each component. The same spectral window was always used. A Gaussian shape was assumed for all Raman lines because of the amorphous state of examined materials (a Lorentzian shape is used for crystalline phase components).

The analysis of multi-spectra sets (images) was performed using in house-produced « PARADIS » software based on the Levenberg-Marquardt method (12,14). The spectra did not receive any 'manual' treatment since the software automatically sets the baseline (for complex backgrounds we introduced Gaussians). A reference spectrum is decomposed first (shape and intensity modelling) and the result is compared with that obtained with Origin® Software. Very similar data are obtained. The result serves as a starting point for one of the neighbouring spectra. Each spectrum is fitted using the data coming from its neighbour, which minimises the standard deviation when decomposition starts. When the model contains only Gaussians and/or Lorentzians, the calculation time is short (a few seconds per spectrum for a ~1 MHz clock PC). The images of any pertinent parameters, such as the wavenumber or the peak intensity, are finally treated using Origin® 5.0 software.

# 3. RESULTS AND DISCUSSION

## 3.1. Microstructure

*Single crystal seeds*

Optical (Fig.1a) and SEM (Fig. 1b) photographs show the PMN-PT cubic crystals of ~50x50x50 $\mu m^3$ that were prepared and selected as templates. Typical green (Fig. 1c) and fired ceramic microstructures (Fig. 2) are shown.

*Oriented green tape*

HTGG process requires an ultrafine matrix, large templates and an appropriate liquid phase for the development of the microstructure. The higher the surface area of the powder, the higher the driving force for the densification. The PMN-PT perovskite powder used exhibits a specific surface area of ~12-15 $m^2/g$ and TEM photograph shows aggregates of about 1 $\mu m$ constituted by nano-particles of 10 to 100 nm. Green tapes of 100-150 $\mu m$ thickness obtained by tape casting exhibit cubic templates oriented with the <001> plane perpendicular to the tape surface (Fig. 1c).

We will consider two types of samples, representative of the most different cases encountered in the synthesis campaigns:

- i) a sample sintered at 1150°C (here after called *#1-sample*), with large crystals dispersed in the ceramic matrix. Apart from large cubic-shaped crystals, many small crystals (in white on optical micrographs) < 10 μm in size are observed (Fig. 2a). The matrix grain size is ~2 μm (Fig. 2e), much smaller than the size of the pristine crystal seeds. The density is 7.66 (i.e. 95% of the theoretical density).

- ii) a sample sintered at 1200°C (*#2-sample*), exhibiting a nearly complete pavement of large sized crystals. The residual matrix is only present along part of the grain boundary network.

These two samples have been prepared using the same seed and nano-powder batches and they are representative of the microstructures obtained in our numerous syntheses. A comparison of corresponding X-ray powder patterns is given in Fig. 3. The density is 7.05 (i.e. 88% of the theoretical density).

In both as-sintered samples the growth area around the pristine crystal seed is obvious: the crystal core appears in black and the periphery in white (Fig.2a,b). Laping and polishing #2-samples made it possible to analyse their evolution with thermal treatment: at the beginning of the thermal treatment well-faceted crystal are observed. Residual smooth pores then appear between adjacent crystals (Fig. 2f).

*X-Ray Diffraction spectra of Tape Cast Textured Ceramics (TCTC)*

XRD spectra recorded on as-sintered surfaces (Fig. 3) show the ceramic sintered at 1100°C presents a random orientation (highest intensity observed for the (100) peak) as for a green tape. A strong increase of the (00λ) diffraction peak intensities was observed for samples sintered at 1150°C, indicating an increase in the crystallographic (001) orientation of the TCTC samples. For ceramics sintered at 1200°C during 10h, the XRD spectrum presents only the peaks due to the (00λ) orientation (Fig 2c). TCTC disks (Φ~14 mm) exhibiting a texture of 0.7 and 0.9 (Lotgering factor) were obtained for ceramics sintered for 10h at 1150 and 1200°C respectively. Examination of the #1-sample microstructure shows crystals with sharp corners and straight edges, regardless of the size, indicating that an equilibrium state similar to the one observed in ref 8 has been reached. On the other hand, #2-samples show smoothed edges and spherical pores (Fig. 2d) which are typical features of ceramics densified by liquid phase sintering with the high vapour pressure of PbO, as could be expected for the higher sintering temperature. No traces of pyrochlore phases were observed.

These results seem to indicate that the texture starts to develop at 1150°C and grows at higher temperature and/or for long sintering. At room temperature, the doublet observed for the

(002) reflection peak reveals the mixture of rhombohedral and tetragonal phases. Above the Curie temperature (200°C), only one peak is observed on the type #2-sample spectra while the doublet remains for the type #1-sample (13). This doublet was thus assigned to two different compositions for the #1 sample in the cubic phase. The most important peak observed at the same position as for the #2-sample (unit-cell parameter a = 4.026 Å) obviously corresponds to the PMN-PT x = 0.35 matrix composition while the second one (4.0185 Å) is related to the x= 0.25 seed composition.

*Phase diagram and growth mechanism*s: Typical DTA/TG results are compared in Fig. 4. The DTA diagram of PbTiO$_3$ single crystals shows a strong endothermic peak at 1295°C, which corresponds to the melting temperature; large hysteresis is observed on cooling with an exothermic crystallisation peak at 1225°C. Note also the pre-transitional asymmetry of the melting endothermic peak. Very similar features are observed for the PMN$_{0.65}$PT$_{0.35}$ single crystals and a random ceramic (sintered from already formed PMN-PT perovskite powders). However, an endothermic jump is observed on heating at ~1180°C in DTA for the ceramic sample while only a smooth pre-transitional effect is observed for the reference single crystals. The strong and narrow endothermic peak observed at 1140°C for PZN-PT single crystals was attributed by Doug and Ye (16) to the partial decomposition of the perovskite phase, leading to the formation of the pyrochlore phase and the segregation of PbO. Non eutectic solidification of PMN-PT with no segregated PbO was observed near 830°C. Very small endothermic events (at ca. 850°C) can be observed for #2 samples by zooming in the DTA traces (Fig. 4). We observe the superimposition of a 1225°C narrow peak on cooling, as for the PT composition. The following equation can be proposed:

$$PMN_{1-x}\text{-}PT_x \implies PMN_{1-x}\text{-}PT_{x-y} + y\ PT \qquad (3)$$

The accident near 1180°C could be due to either i) a strong sintering having the effect of decreasing the contact area between the sample and the crucible or ii) a second order (crystalline phase) or Tg (glassy phase) transition or iii) a very broad "incongruent" melting endothermic effect. Because the phenomenon is reversible, a second order phase transition appears to be a reasonable explanation. However, we see a broad exothermic peak on cooling from ~1300 to 1180°C, which could correspond to the solidification of a "glassy" phase mixing non-stoichiometric B/B' domains. Note the kink observed at ~1000°C simultaneously with the onset of weight loss is certainly due to the sample sintering and subsequent decrease of the sample-crucible contact. The second kink observed on the TG plot corresponds to the melting point and the increase of the PbO evaporation rate. Details of the DTA runs performed successively on the two types of textured ceramics show marked differences: for the #1-sample first heating cycle, a rather large endothermic anomaly is observed between 1000 and 1100°C as already observed on a random ceramic. A plateau is then observed up to ~1240°C, i.e. up to the onset of melting. On the contrary the #2-textured ceramic shows a small exothermic bump. These different behaviours confirm that growth conditions are different for the two types of samples. It is worth noting that no pyrochlore phase was observed in the two textured ceramics either on XRD spectra or on Raman spectra. The schematic phase diagram proposed in Fig 5 is different to that proposed for the homologous Zn-solid solution (16). We postulate the existence of two liquid phase regions to explain the observed grain shape and DTA traces. By analogy with the PZNT system (16), a eutectic composite is expected close to the 75 % PbO composition. #1-samples are prepared at 1150°C, i.e. at the temperature of the endothermic anomaly (region I) whereas #2-samples are obtained in the region II plateau, just below the melting point. Growth conditions in this later region are very sensitive to the PbO volatility and to any modification of the PMN-PT melting temperature. Any local heterogeneity of the solid solution will modify the liquid (composition, viscosity, …). The existence of two different liquid phases appears the most reasonable explanation for the observed microstructure: the first one would appear at ~1185°C, enabling for the initial growth of well-faceted crystals on cubic seeds and the second one would appear above 1250°C (see the shape of the strong endothermic events on Fig. 4) concomitant with a stronger PbO evaporation and thus a composition shift within the solid solution range.

Spherical pores at the grain boundary relic appear characteristic of #2-samples (Fig. 2f). Different mechanisms could be at their origin: i) high PbO partial pressure or ii) the Kirkendal effect (17). This second mechanism would imply that vacancies and ions ($Mg^{2+}$ or $Pb^{2+}$ ions?) diffuse simultaneously but at different rates. This would be consistent with the modification of the short-range composition, as schematized by equation 3. More data are needed to further understand this last phenomenon and its relationship with B/B'ordered domains.

### 3.2. Raman signature and composition dependence

Fig. 6 compares the Raman spectra of different matrices, from pure $Pb(Mg_{1/3}Nb_{2/3})O_3$ (x = 0) to pure $PbTiO_3$ (x = 1). The matrix grain size and the grain size far from the seeded regions is ~ 0.5 μm, much smaller than the size of the crystal seeds. This very small grain size makes, that even with high magnification objectives, the spectra recorded are "powder" Raman spectra taking the different polarisations into account.

According to the relaxor structure, Raman peaks are very broad. This has already been related to short-range intrinsic disorder (18-24), i.e. there are zones where the cations are displaced from their high-symmetry positions giving rise to a local electric dipole. Because Raman spectroscopy is sensitive both to the atomic dynamics (through the wavenumber centre of gravity) and the local electric charge transfer (through the intensity), local electric defects disturb the spectrum a lot (25).

The Raman spectrum can be separated into three regions, the low-wavenumber region (<150 cm$^{-1}$), the mid-wavenumber region (150-380 cm$^{-1}$) and the high-wavenumber region (> 380 cm$^{-1}$). According to previous literature reports, the low-wavenumber region is attributed to vibrations associated with motions of A cations (accordingly, corresponding peaks are very narrow in peroskite-type structures free of any A cation disorder). We chose to compare our samples spectra with the $Ba_6Nb_2O_{11}$ compound (26) because it is an "ordered" Nb-based perovskite free of A cation disorder (very narrow low wavenumber components are observed). The mid-wavenumber region involves bending motions of the $BO_6$ octahedron and the high-wavenumber region the stretching modes. These high-wavenumber modes are dominated by vibration of $O^{2-}$ anions involving the most rigid cation-oxygen bonds inside the $BO_6$ octahedron (breathing modes). As expected, Ba/M substitution (giving fast oxygen/proton diffusion) broadens this region drastically because different short-range order/domains (26). Comparison of the different spectra shows that the 700-900 cm$^{-1}$ peak is dominated by the Nb-O stretching components (Fig. 6). On the other hand, the 500-700 cm$^{-1}$ region mainly reflects the contribution of Ti-O modes, their lower intensity arising from the lower atomic number and, hence, a smaller number of electrons involved in the bond. The merging of the different well-defined peaks of a single B site occupied Ti-perovskite into a broad pattern for the mixed one is assigned to the special short-range disorder of the relaxor structure. Note that the centre of gravity of the band is very similar to that of the $Ba_6Nb_2O_{11}$ perovskite, in agreement with a dominant contribution of modes involving Nb-containing vibrational entities. Thus, in a first approximation based on the quasi-molecular description, we can say that analysis of the 700-900 cm$^{-1}$ region will give information on the local structure of the $NbO_6$ octahedron. The centre of gravity and area in this band could be used as a probe to follow compositional changes, if it is assumed that the structure remains unchanged or that its changes are averaged by the band broadness. On the other hand, the behaviour of the 200-400 cm$^{-1}$ region will give information on the short-range arrangement of the A cation–$BO_6$ octahedron. Note the ca. 100 cm$^{-1}$ narrow peak, characteristic of the long range A order, drastically broadens with Nb/Mg substitution as previously observed for parent perovskites (26).

*Composition dependence*

All spectra have been decomposed as shown in Figure 6. Because of the high short-range disorder and non-resonant character of Raman scattering, each component was considered as a Gaussian (Table 1). An example of a component area plot as a function of x composition is given in Fig. 7. The plot of the centre of gravity of the main component (Fig. 8) shows a slow but regular shift with x composition for the 580, 750 and 800 $cm^{-1}$ peaks.

The main changes associated with the PMN/PT solid solution consist in the increase of the splitting of the $BO_6$ stretching modes (750 and 805 $cm^{-1}$ components, Table 1) and the down-shift of the peak at ca. 600 $cm^{-1}$ (involving Ti-O stretching modes, Fig. 8 and Table 2). Note also that Mg/Nb substitution and PMN/PT phase mixing lead to an increase of the low-wavenumber band intensity, as observed on heating ion-conducting perovskites (26). Conversely, this effect is reduced under pressure (20,21). We can expect the unit-cell expansion observed on heating or by decreasing pressure to promote stronger dipoles and the intensity of the low-wavenumber region can probably be used as a parameter to measure the relaxor degree.

*Sensitivity and experimental error*

Spectroscopic changes are rather small. In order to assess the sensitivity of the measurement, we performed different line scans across selected (large) single crystals, as shown in Fig. 1a. Wavenumber accuracy is determined by comparing the data of about 20 line scans (Table 1). The error (~1 $cm^{-1}$) corresponds to the expected value for our experimental conditions (fixed position of the two-stage monochromator with ~1 pixel per $cm^{-1}$). Reproducibility of the monochromator position is controlled using an Hg lamp and $CCl_4$ reference. The peak area appears to be very sensitive to the pristine seed growth process: an example is shown in Fig. 9 for the wavenumber of the ca. 750 $cm^{-1}$ component. Spectra were recorded along a line from the centre of the crystal seed to the periphery. The peak area – and wavenumber– is stable up to 20 μm from the crystal centre and then drops from there to the periphery. This fact reveals growth irregularity assigned to a composition change at the end of the flux growth process.

Table 2 summarizes the data obtained for the different matrix compositions. Taking Table 1 data into account and assuming that the observed variation must be at least 3 times larger than the mean dispersion to be considered significant, we can conclude that the components at ca. 430, 750 and 800 $cm^{-1}$ are sensitive probes.

*Phase characterisation*: Raman spectroscopy has intrinsic selectivity due to that wavenumber position arises mainly from the mechanics of atoms although the peak intensity arises from the local conductivity. The wavenumber set chiefly depends on the atoms structure (mass and bond length). On the other hand, the local charge transfer and the bond polarisability determine the band intensity. Consequently the plot of the wavenumber as a function of x-composition (Fig. 8) obviously shows the homogeneous behaviour expected for a solid solution. On the contrary, peak area plot versus composition shows different regimes which can be related to the different unit-cell distortions (with different ferroelectric properties) shown by X-ray diffraction (13,27): a single rhombohedral cell (R) for x<0.25, a mixture of two monoclinic cells ($M_1 + M_2$) between 0.25 and 0.35 and a mixture of monoclinic and tetragonal cells ($M_2 + T$) for x> 0.35. Note the large electric disorder at the very local scale breaks the periodicity and the Raman spectrum reflects the vibrational density-of-state of the entire Brillouin zone. Polarization effects are thus small and this makes our assumption to consider the crystal spectrum as a "powder" spectrum reasonable.

### 3.3. Raman Imaging

Raman spectra and their direct and "smart" images can be used to understand and even predict some properties (12). Figure 10 shows a direct Raman image recorded on an as-sintered #1-sample. The peak area measured in a given wavenumber window is plotted for each point of measure (3000 points). High values are observed at the periphery of some grown grains. Corresponding spectra show a very large scattering ascribed to fluorescence (Fig. 10). Two types of fluorescence can be observed in our ceramics. The first originates from 3d or 4f doping ions or impurities (this undesirable doping can be due to rare earth and transition element traces in the starting powders (28,29). The second originates from organic bonds grafted at the pores' surface; this type of fluorescence is encountered in ancient or excavated ceramics or on polished porous samples (polishing agent residues). It is easily eliminated by thermal treatment or by using high laser power illumination and/or blue/violet excitation and can consequently be ruled out for the samples sintered at high temperature and not polished. The PMN-PT matrix powder was doped with Mn, the fluorescence thus being assigned to the presence of specific 3d/4f elements (Mn element and their impurities) concentrated at the periphery of larger crystals. This is consistent with a concentration of doping impurities in the material obtained by final solidification from the mother interstitial liquid phase. Wallace et al. (30) claimed that cubic-shaped crystals formation is related to a large liquid phase volume fraction (>0.4). The lack or the very small occurrence of PbO/eutectic melting events shown byon DTA traces demonstrates that the liquid phase volume fraction is not the prominent parameter.

In order to distinguish the Raman spectrum from the fluorescence background, we describe the later phenomena with a broad Gaussian band centered in the analysed window. Results are shown in Fig. 11. This example shows a typical spectrum after the fluorescence contribution has been removed, to be compared with the Fig. 12-spectrum whose the data are free of fluorescence).

*#1-Sample*:

The images showing the calculated centre of gravity and peak area clearly show three regions (Fig. 11): the matrix, the grains' core (i.e. single crystal seeds) and their periphery. Clear-cut images are obtained for the main (ca. 800 cm$^{-1}$) component. The difference in the peak area of the main matrix and the grain core is straightforward: the highest values reveal the pristine seed surface (ca. 50 x 50 $\mu m^2$, as in Fig. 1). The growth mechanism seems to be limited to the interface between the seed and the matrix. The seed surface appears not to be modified. Mean values calculated for each region are summarized in Table 3. Note the size of the core region corresponds well to the surface of pristine seeds. The very high peak area measured on the seed surface indicates that the surface was preserved from any modification. This large contrast also results from the difference in ferroelectric behaviour.

*#2-Sample*

The images show very good homogeneity (Fig. 12): whereas the wavenumber and peak area shifts are ca. 30 cm$^{-1}$ and 3800 counts·cm$^{-1}$ in the type #1-sample (Table 4), these shifts are only 6 cm$^{-1}$ and 200 counts in the type #2-sample. Only very few locations show out-of-range values. They correspond to the centre of rare relics of pristine crystalline seeds. It is clear that the surface of the pristine seeds has been modified. Typically, the observed crystal size is 150 x 150 $\mu m^2$. Very similar results are obtained for the different components (Table 3). The wavenumber and peak area of the contour are very similar to that of the matrix. This indicates that the composition of the material that grows at the seed-matrix interface is determined by the matrix composition and that it is therefore not necessary that the seed composition exactly correspond to the desired final composition. Furthermore the low content of the initially dispersed seeds (a few % in weight) makes the contribution of the matrix dominant. These data are consistent with X-ray diffraction patterns, representative of average structure (and long distance order). This reveals only one texture for the #2-sample, the composition of which is close to that of the ceramic matrix, but two different composition textures for #1-samples.

The local composition has been calculated using the correlation established in the first part of the work. According to the low dispersion of the data in the type #2 sample, the composition is rather homogeneous: the grain core x composition lies between 0.25-0.4 as a function of the seed and the matrix composition is close to 0.3-0.35.

## 4. CONCLUSION

Textured ceramics corresponding to Pb(Mg$_{1/3}$Nb$_{2/3}$)O$_3$-PbTiO$_3$ (PMN$_{(1-x)}$-PT$_x$) solid solutions have been prepared by homo-epitaxial templated grain growth (HTGG), using PMN-PT cubic templates and nanoparticles and controlled atmosphere sintering. The wavenumber centre of gravity versus x composition relationship enables the x composition to be extracted and, hence, for its distribution to be imaged in the pristine seed and between the seeds and the pristine ceramic matrix. Ceramics displaying medium and high texturing were obtained by different growth conditions: at low temperature, sharp cornered and straight edged single crystal seeds develop through the liquid phase at the crystal/matrix interface. It is clear that the crystalline material growing from an interstitial liquid phase at the seed-matrix interface concentrates the impurities soluble in the liquid phase and that the composition of the grown perovskite is determined by the matrix composition. It is thus not necessary that the composition of the pristine seed dispersed in the green matrix has the target composition. The different compositions shown by Raman imaging are consistent with the two textures observed by X-ray diffraction. Two different growth mechanisms have been identified and a schematic pseudo-binary phase diagram is proposed. At high temperature, the matrix is consumed and crystals develop in all the ceramics, with the help of a liquid phase I (< 1185°C) or a liquid phase II (> 1200°C). High temperature processed materials have a secondary closed porosity (bubbles) along grain boundary phantoms. Smart Raman imaging shows that the final composition is very close that of the matrix. The use of an appropriate procedure and software to model the Raman fingerprint and extract parameters is mandatory to clear the spectra from undesired contributions (for instance the fluorescence, observed in many cases). Useful information can be extracted in our case in spite of the broadness of Raman patterns after determination of the method accuracy. Main conclusions are summarized in Table 4. The bands' centre of gravity reflects the homogeneous character of the PMN-PT solid solution, while their peak area shows the local B/B' heterogeneity and associated unit-cell distortions. This first use of Raman imaging of textured ceramics deserves new analyses in combination with other techniques (EDAX, EPMA, etc.). The main advantage of the non-destructive Raman technique is its potential to image composition shift and unit-cell distortion without a complex preparation of the samples.

**ACKNOWLEDGEMENTS**

This work was supported in part by "PYRAMID" project under European Fifth Framework Programme. The authors thank Mrs Annie Marx for technical support in ceramic processing, Mrs Fatima Ammamou and Mr Gérard Sagon for helping with recording of the spectra and Mrs Anne-Marie Lagarde for the figures.

**FIGURE CAPTIONS**

**Figure 1**: (a) Optical and (b) SEM photographs of PMN-PT cubic templates; (c) optical microphotograph of PMN-PT green cast tape showing typical seed distribution. Vertical and horizontal lines show the path analysed by Raman scattering across a $PMN_{0.75}PT_{0.25}$ single crystal.

**Figure 2**: Optical microphotographs recorded on as-sintered surfaces of (a) representative #1- and (b) #2-samples; (c-d and f) details recorded on polished Tape Cast Textured Ceramic (TCTC) surface, showing grain boundaries and pore network displaying the texture fraction of 0.7 (left) and 0.9 (right); (e) the matrix intergranular fracture micrograph shows the micronic grain size.

**Figure 3**: Representative X-ray diffractogramms of homoepitaxial template grain growth (HTGG) Tape Cast $PMN_{1-x}PT_x$ green tape *and* ceramics sintered at 1100°C (random ceramic), 1150°C (f= 0.77, #1-sample) and 1200°C (f=0.9, #2-sample).

**Figure 4**: DTA traces recorded on random PMN-PT ceramic, #1- and #2-textured ceramic pieces (x = 0.35, details are given on the right side: 1st cycle up to 1180°C, 2nd cycle up to 1250°C and 3rd cycle up to 1280°C). A comparison is made with pure PT single crystals.

**Figure 5**: Schematic of the pseudo-binary $PMN_{1-x}PT_x$ phase diagram showing the growth regions for #1 and #2 samples.

**Figure 6**: Representative Raman spectra of PT, $PMN_{1-x}PT_x$ and pure niobate perovskites. Detail of the peak fitting is given for the 400-950 cm$^{-1}$ range. * , plasma line.

**Figure 7**: Plots of the relative area of the main components at ca. 430, 580, 750 and 800 cm$^{-1}$, as a function of x composition for PMN$_{1-x}$PT$_x$ ceramics. Typical error bars are given. Lines are guide for the eyes.

**Figure 8**: Plot of the component wavenumbers as a function of x-composition. Calculated error bars are given.

**Figure 9:** Examples of horizontal line scans recorded from the centre to the periphery of a crystal seed (Fig. 1a);top: area of the ca. 750 cm$^{-1}$ component), bottom centre of gravity.

**Figure 10**: Example of direct image obtained using Raman intensity in a given wavenumber window (here 200-1200 cm$^{-1}$). Mapped area 700x208 μm; 30x10 spectra, objective: x10, λ = 632 nm, see Fig. 8 for the optical micrograph. High intensity regions appear in white/grey, low intensity regions in black. An example of high-intensity spectrum is shown.

**Figure 11**:  #1-sample**:** mapped area 700x208 μm ; 30x10 spectra, objective: x10, λ = 632 nm; vertical scale is multiplied by ~3. Wavenumbers (left) and peak area (right) are mapped for the two components of the main Raman peak (see Fig. 4). A typical point-spectrum is shown after subtraction of the fluorescence "background". Mean values are given for each zone: matrix, crystal grain core and contour.

**Figure 12**:  #2-sample: mapped area: 650x600 μm, 65x60 spectra; objective: x10, λ = 514 nm. Wavenumbers (left) and peak area (right) are mapped for the two components of the main Raman peak (see Fig. 4). A typical point-spectrum is shown. Mean values are given for each zone; matrix, crystal grain core and contour.

**Table 1 :** Line scan mean values and their dispersion
measured on seed crystals

| <Wavenumber> | <Peak Area> |
|:---:|:---:|
| cm$^{-1}$ | % |
| 437.2 ± 0.6 | 1 ± 0.1 |
| 509.3 ± 1.7 | 8.4 ± 1 |
| 580.8 ± 0.7 | 28.6 ± 1.4 |
| 751.7 ± 1.2 | 43.5 ± 3.1 |
| 806.1 ± 0.7 | 18.5 ± 2.9 |

**Table 2** : Comparison between the wavenumber (ν) and peak area (A) ; mean data dispersion (<Δν>) is given for the main components for the different compositions and the observed shift assumed to be significant if > 3 x (maximal) mean.

| ν / cm$^{-1}$ | 430 | 500 | 580 | 750 | 800 |
|---|---|---|---|---|---|
| x PT | | | <Δν> | | |
| 0.2 | 0.6 | 1.8 | 1.3 | 0.3 | 0.4 |
| 0.25 | 0.4 | 0.8 | 0.6 | 0.3 | 0.2 |
| 0.35 | 0.4 | 0.7 | 1.1 | 0.6 | 0.3 |
| ν$_{0.35}$ −ν$_{0.2}$ cm$^{-1}$ | 5 | - 4.6 | - 3.2 | - 20.4 | 10.8 |
| Validity | Yes | No | No | Yes | Yes |
| Rate cm$^{-1}$/% | 0.3 | | | - 1.4 | 0.7 |

| A / % | 430 | 500 | 580 | 750 | 800 |
|---|---|---|---|---|---|
| x PT | | | <Δν> | | |
| 0.2 | 0.1 | | 0.5 | 0.6 | 0.2 |
| 0.25 | 0.1 | | 0.5 | 0.7 | 0.2 |
| 0.35 | 0.1 | | 0.3 | 0.7 | 0.6 |
| ν$_{0.35}$ −ν$_{0.2}$ | 0.5 | | 6.5 | -14 | 9.9 |
| Validity | Yes | No | Yes | Yes | Yes |
| Rate %/% | | | 0.4 | 0.9 | 0.7 |

**Table 3**: Comparison of the mean characteristic values

| Reference | cm$^{-1}$ | | #1-Sample cm$^{-1}$ | #2-Sample cm$^{-1}$ |
|---|---|---|---|---|
| Seed[a] | $\langle v \rangle = 806$ | Core | $\langle v \rangle = 833$[b] | $\langle v \rangle = 800$ |
| (25% PT) | $\langle v \rangle = 752$ | | $\langle v \rangle = 690$ | $\langle v \rangle = 744$ |
| | | Contour | $\langle v \rangle = 813$ | |
| | | | $\langle v \rangle = 715$ | |
| Matrix | $\langle v \rangle = 811$ | Matrice | $\langle v \rangle = 804$ | $\langle v \rangle = 806$ |
| (35% PT) | $\langle v \rangle = 750$ | | $\langle v \rangle = 751$ | $\langle v \rangle = 750$ |

[a]: Seed spectra are polarised

[b]: highest observed value

**Table 4**: Information extracted by Raman Imaging

| Question | Raman signature | Information |
|---|---|---|
| PMN/PT mole ratio | Peak wavenumber | Homogeneity |
| Unit-cell distortion | Peak area | Structure |
| Active interface | Fluorescence | Microstructure stability |

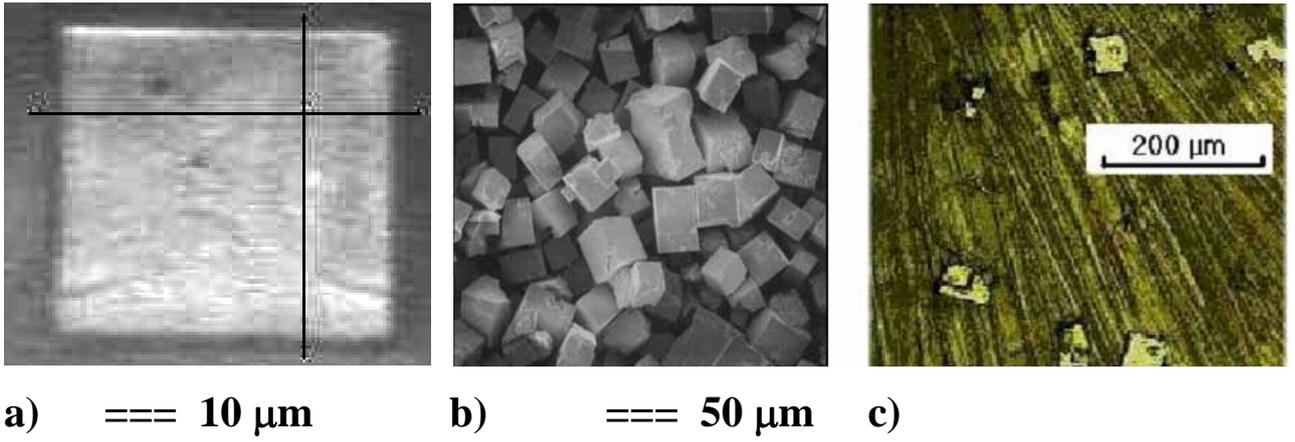

**a)**     **=== 10 µm**       **b)**       **=== 50 µm**     **c)**

**Figure 1**

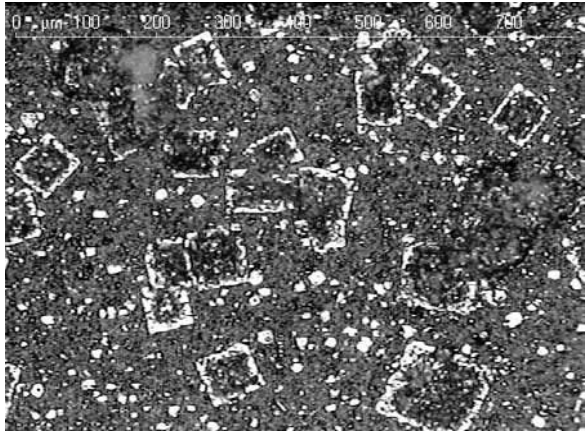 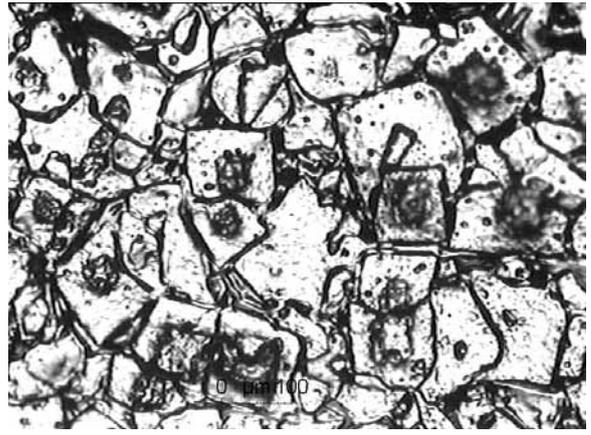

**a)**                     == 50 μm  === 100 μm                          **b)**

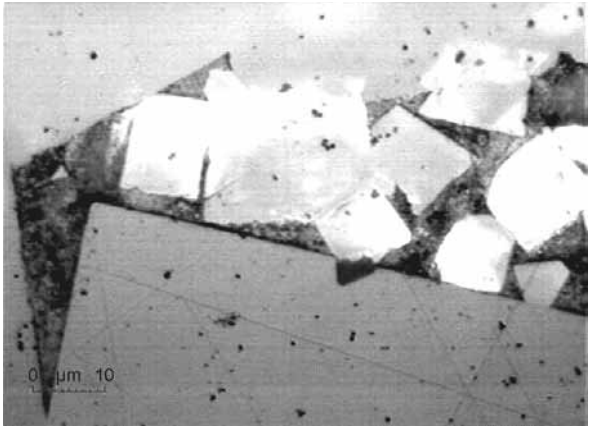 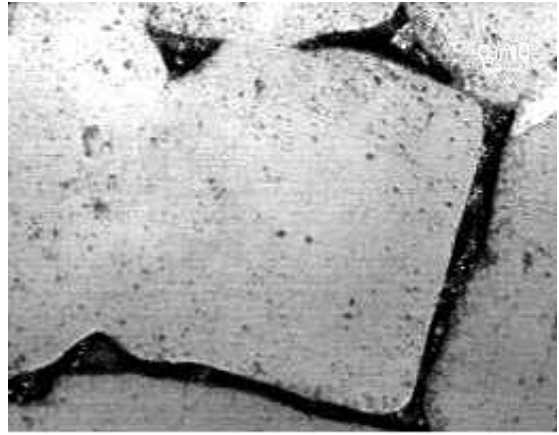

**c)**                     === 10 μm  == 10 μm                          **d)**

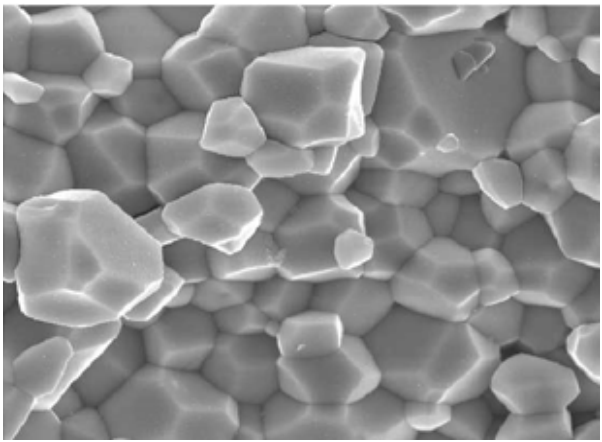 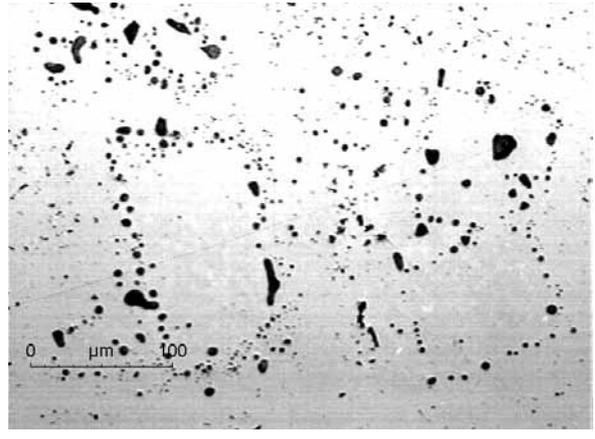

**e)**                     === 2 μm  ====== 100 μm                          **f)**

**Figure 2**

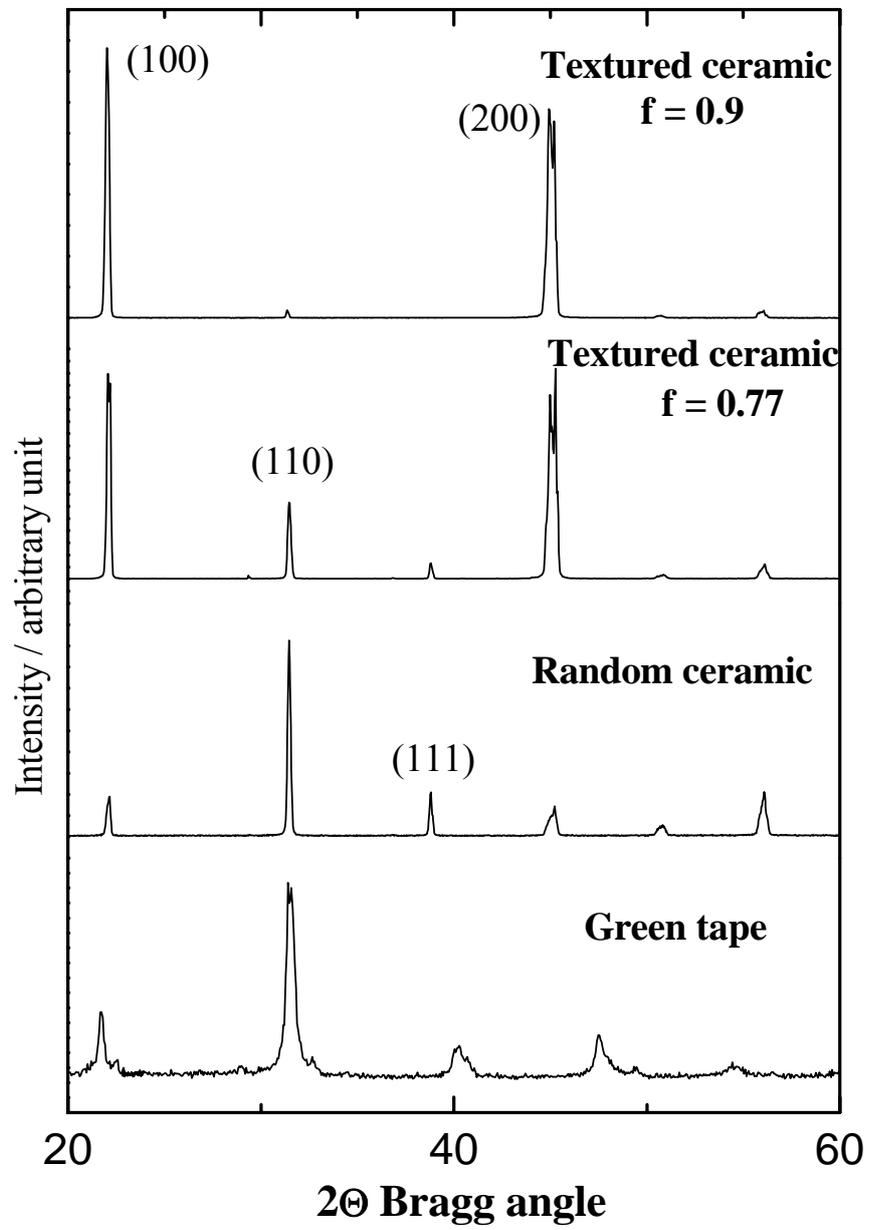

**Figure 3**

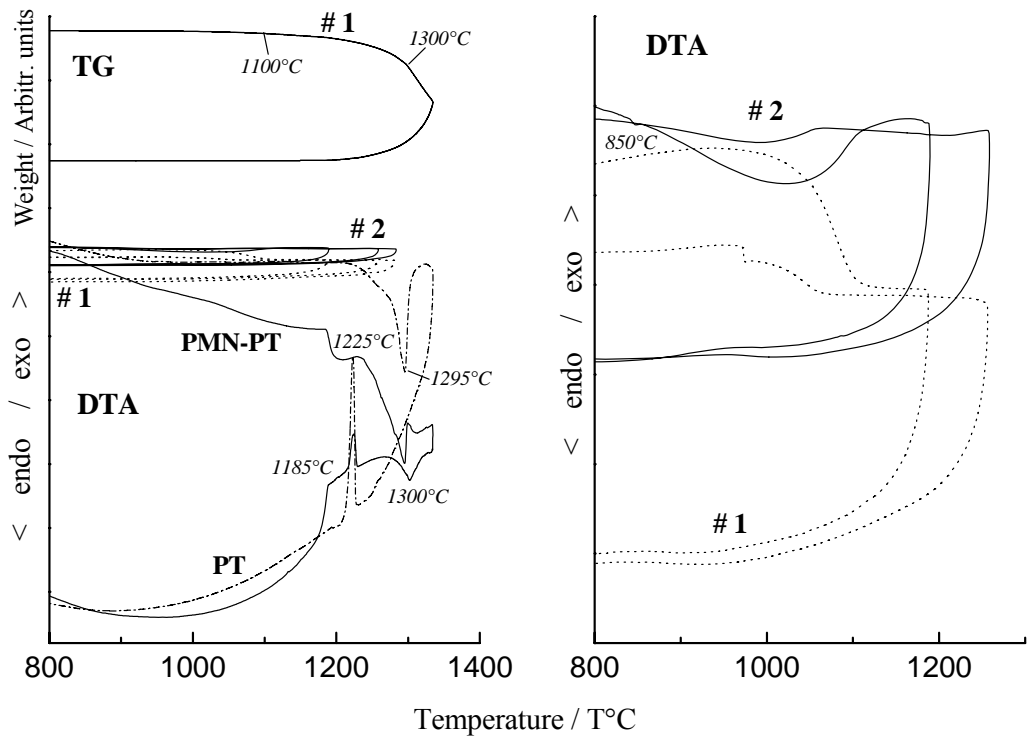

**Figure 4**

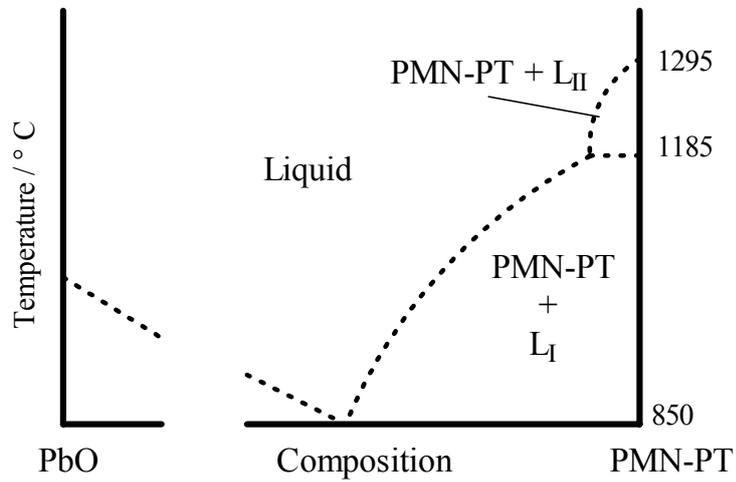

**Figure 5**

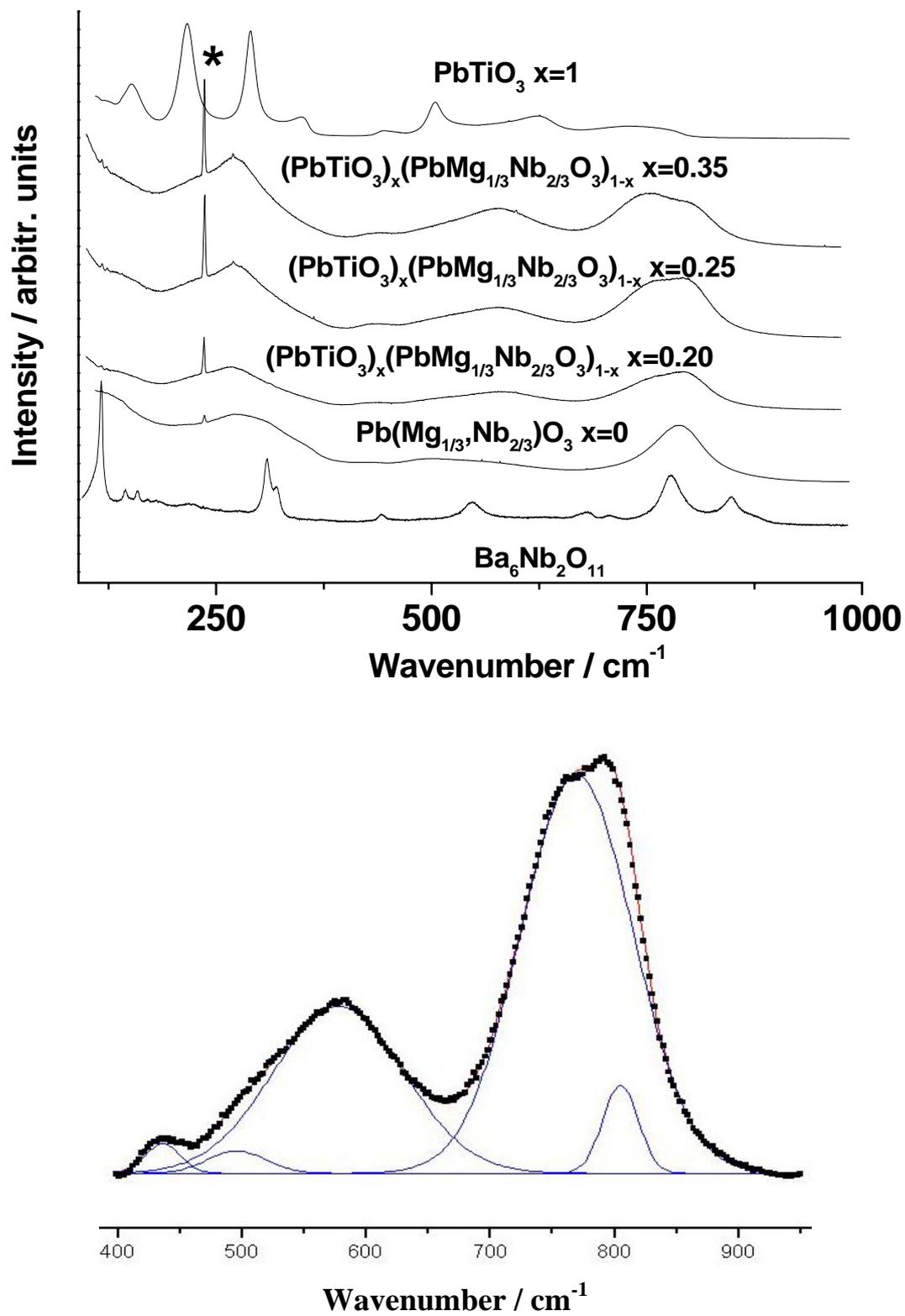

**Figure 6**

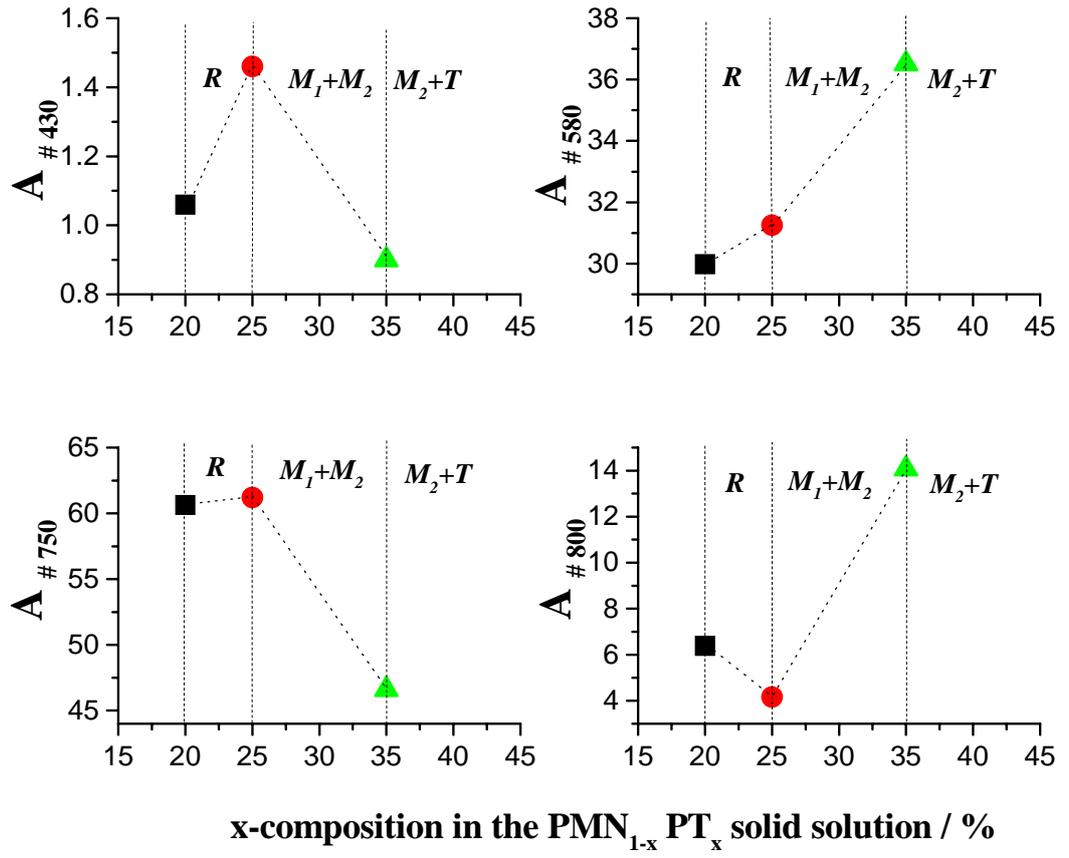

**x-composition in the PMN$_{1-x}$ PT$_x$ solid solution / %**

**Figure 7**

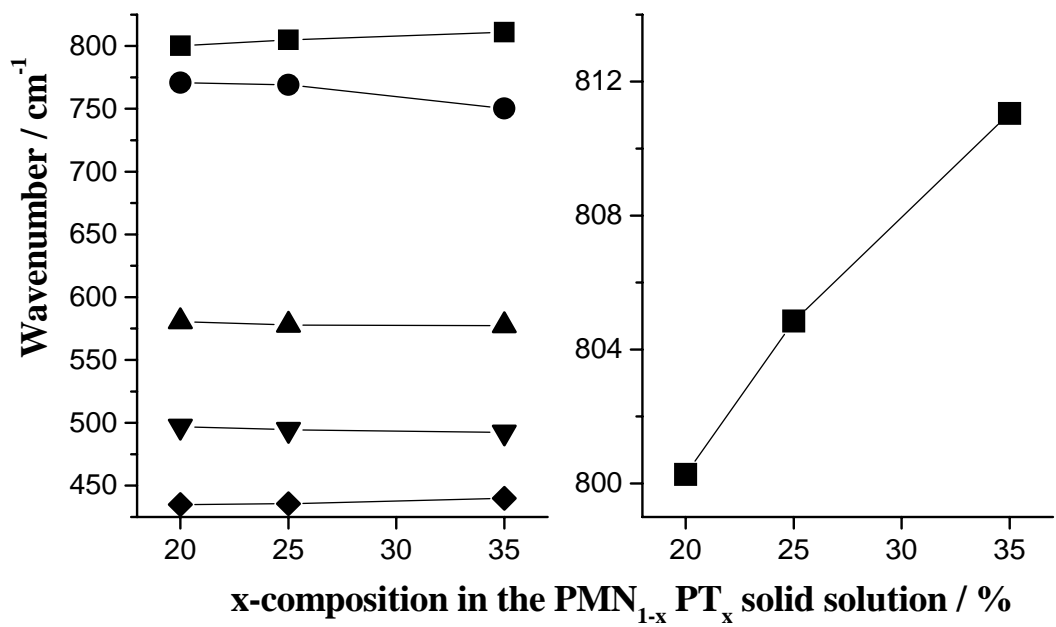

Figure 8

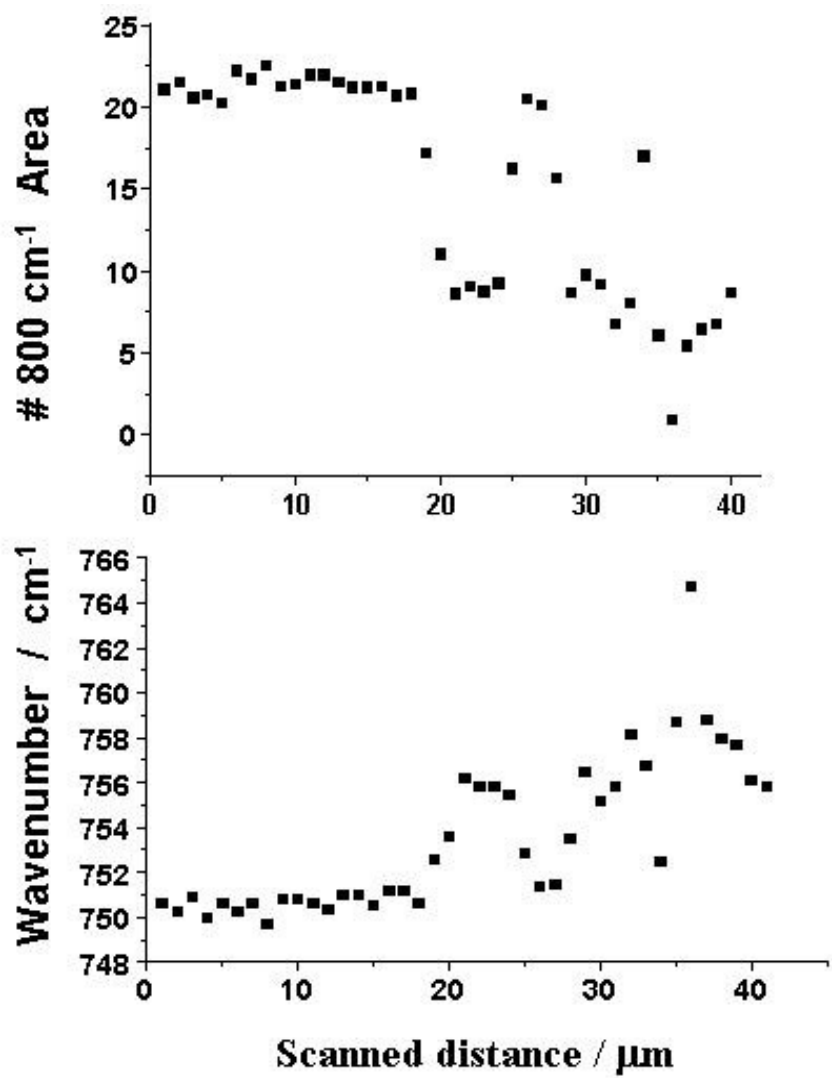

**Figure 9**

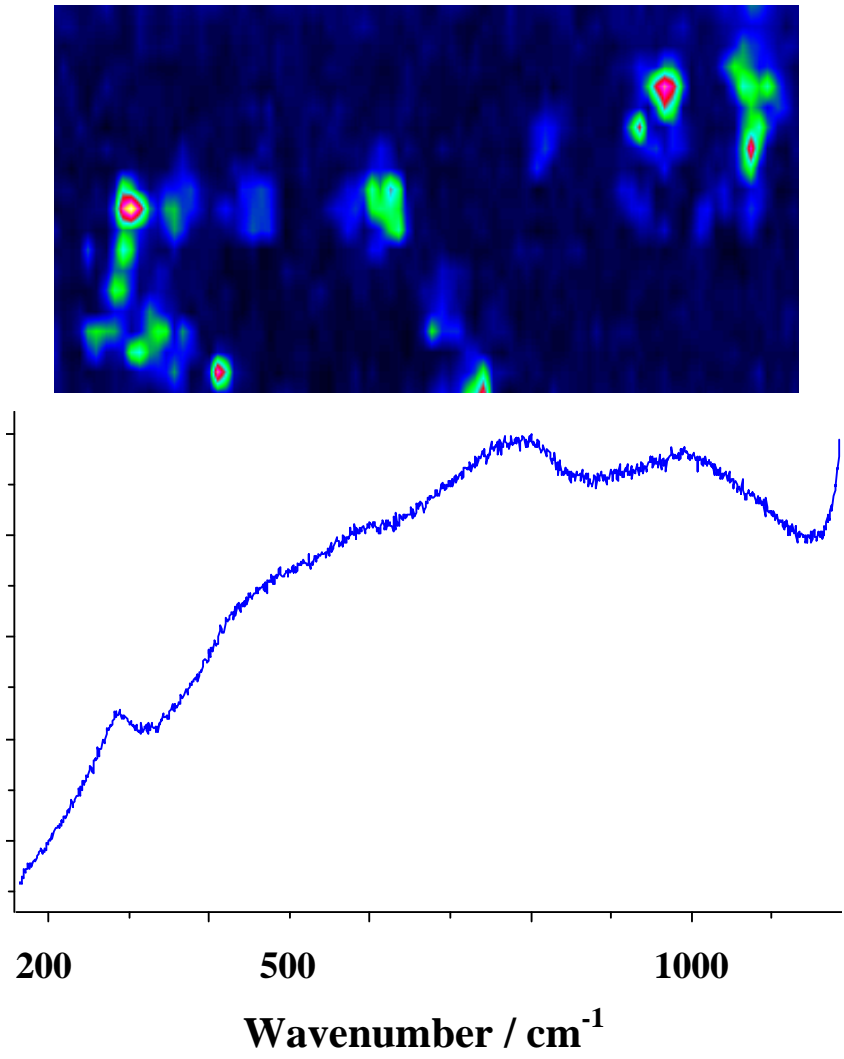

**Wavenumber / cm⁻¹**

**Figure 10**

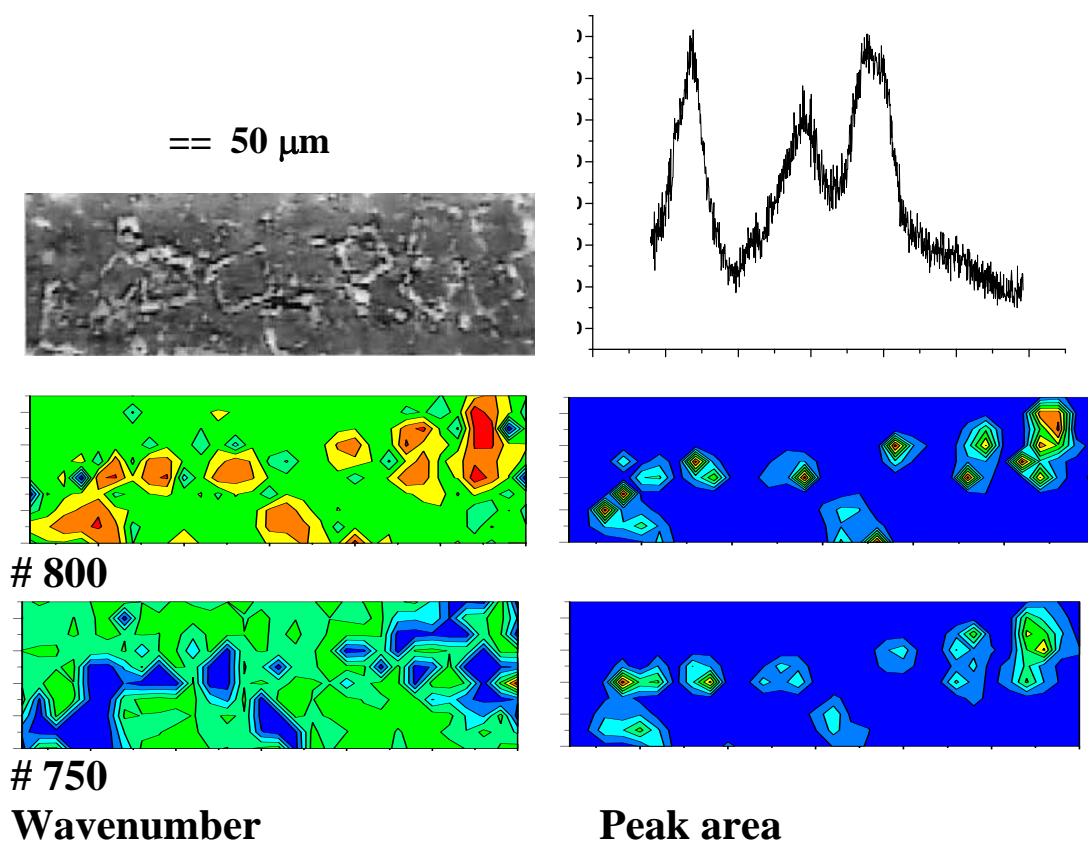

== 50 μm

**# 800**

**# 750**

**Wavenumber**          **Peak area**

**Figure 11**

=== **50 μm**

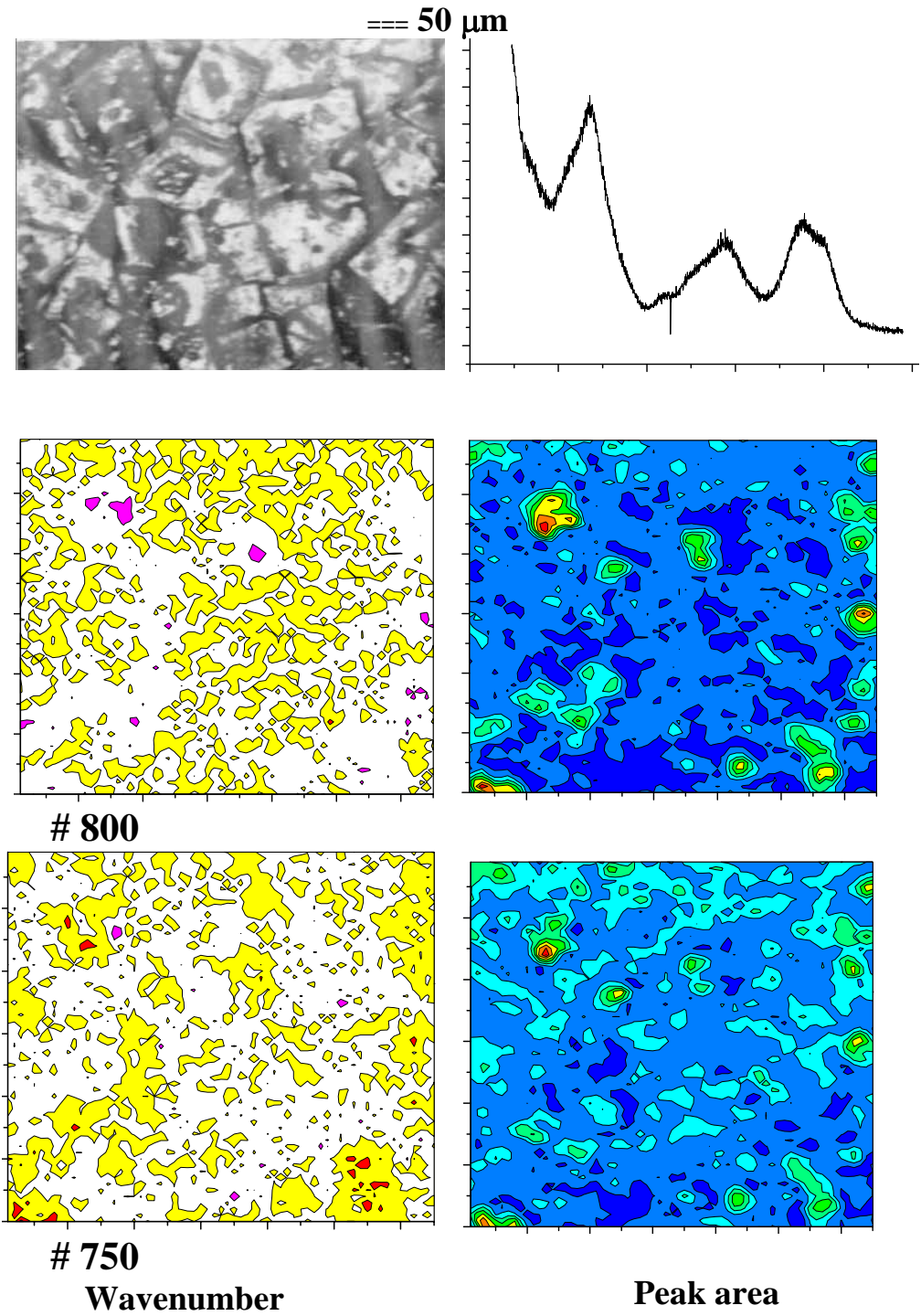

# 800

# 750
Wavenumber

Peak area

**Figure 12**